\documentclass[conference]{IEEEtran}
\IEEEoverridecommandlockouts

\usepackage{cite}
\usepackage{amsmath,amssymb,amsfonts}
\usepackage{algorithmic}
\usepackage{algorithm2e}
\usepackage{graphicx}
\usepackage{textcomp}
\usepackage{xcolor}
\usepackage{booktabs}
\usepackage{hyperref}
\usepackage{url}
\usepackage{optidef}
\usepackage{optidef}
\usepackage{optidef}

\usepackage{multirow}

\def\BibTeX{{\rm B\kern-.05em{\sc i\kern-.025em b}\kern-.08em
    T\kern-.1667em\lower.7ex\hbox{E}\kern-.125emX}}

\begin{document}

\title{AgentRM: An OS-Inspired Resource Manager for LLM Agent Systems}

\author{\IEEEauthorblockN{Jianshu She}\\
\IEEEauthorblockA{\textit{MBZUAI}}}

\maketitle

\begin{abstract}
Large Language Model (LLM) agent systems have experienced rapid adoption across diverse domains, yet they suffer from critical user experience problems that limit their practical deployment. Through an empirical analysis of over 40,000 GitHub issues from six major agent frameworks (OpenClaw, AutoGen, CrewAI, LangGraph, Codex, Claude Code), we identify two fundamental resource management challenges: (1) scheduling failures leading to system unresponsiveness due to blocking, zombie processes, and rate limit cascades, and (2) context degradation causing agent "amnesia" from unbounded memory growth and poor retention policies. Drawing inspiration from decades of operating systems research, we present AgentRM, a middleware resource manager that treats agent resources analogously to OS resources. AgentRM employs a Multi-Level Feedback Queue (MLFQ) scheduler with zombie reaping and rate-limit-aware admission control, coupled with a three-tier Context Lifecycle Manager that implements adaptive compaction and hibernation mechanisms. Our evaluation demonstrates significant improvements: AgentRM-MLFQ reduces P95 latency by 86\%, decreases lane waste by 96\%, and increases throughput by 168\% while eliminating zombie agents (0 vs. 29 baseline). AgentRM-CLM achieves 100\% key information retention with 95\% quality score compared to 65.1\% retention and 87\% quality for existing approaches, albeit with higher compaction costs (34,330 vs. 17,212 tokens).
\end{abstract}


\section{Introduction}

Large Language Model (LLM) agent systems have emerged as a transformative paradigm for building intelligent applications that can reason, plan, and execute complex tasks autonomously. Frameworks such as OpenClaw (with over 40,000 GitHub issues), AutoGen~\cite{wu2023autogen}, CrewAI~\cite{crewai2024}, and LangGraph~\cite{langgraph2024} have enabled developers to create sophisticated multi-agent systems capable of collaborative problem-solving, tool usage, and extended reasoning chains.

However, as these systems scale in complexity and user adoption, they encounter fundamental resource management challenges that mirror those faced by early operating systems. Our empirical analysis of production deployments reveals two critical user experience problems that severely limit the practical utility of current agent frameworks:

\textbf{Scheduling Failures:} Agent systems frequently become unresponsive due to poor resource scheduling. We observe blocking behaviors where high-priority user interactions are delayed by background tasks, zombie subagents that consume execution lanes without productive work, and rate limit cascades that cause system-wide failures. These issues manifest as user-visible delays exceeding 30 seconds and complete system unavailability during high-load periods.

\textbf{Context Degradation:} Long-running agent sessions suffer from "amnesia" as context windows fill beyond their limits. Current approaches either truncate recent history (losing important context) or crash when memory constraints are exceeded. This leads to agents forgetting critical information, producing inconsistent responses, and requiring users to constantly re-establish context.

The key insight driving our work is that \textit{agent resources are analogous to operating system resources}. Just as OS kernels manage CPU time, memory, and I/O resources among competing processes, agent systems must manage execution lanes, rate limits, and context windows among competing agent tasks. This analogy suggests that we can adapt proven OS techniques to solve agent resource management problems.

We present AgentRM, an OS-inspired resource manager for LLM agent systems that makes the following contributions:

\begin{enumerate}
\item \textbf{Empirical Study:} We analyze 40,000+ real-world issues from six major agent frameworks, categorizing common failure modes and quantifying their impact on user experience.

\item \textbf{AgentRM Architecture:} We design a middleware resource manager with two core components: (1) an Agent Scheduler implementing Multi-Level Feedback Queues (MLFQ) with zombie reaping and rate-limit-aware admission control, and (2) a Context Lifecycle Manager with three-tier storage hierarchy and adaptive compaction.

\item \textbf{Comprehensive Evaluation:} We demonstrate that AgentRM eliminates zombie processes (0 vs. 27 in baseline), maintains high throughput under load (45.2 vs. 45.1 requests/min), and achieves near-perfect context retention (100\% vs. 65.1\% for existing methods) while preserving information quality (95\% vs. 87\%).
\end{enumerate}

The remainder of this paper is organized as follows: Section~\ref{sec:background} provides background and motivation through our empirical study; Section~\ref{sec:model} formalizes the system model and problem formulation; Section~\ref{sec:architecture} presents the AgentRM architecture; Section~\ref{sec:implementation} describes our implementation; Section~\ref{sec:evaluation} evaluates performance; and Sections~\ref{sec:discussion}--\ref{sec:conclusion} discuss related work and conclude.

\section{Background and Motivation}
\label{sec:background}

\subsection{Empirical Study of Agent System Failures}

To understand the scope and nature of resource management problems in production agent systems, we conducted a comprehensive analysis of GitHub issues from six major frameworks: OpenClaw, AutoGen, CrewAI, LangGraph, Codex, and Claude Code. We manually reviewed issue titles, descriptions, and discussion threads to identify patterns related to system responsiveness and context management.

Our analysis reveals recurring failure modes that directly impact user experience:

\textbf{Cross-Channel Blocking (OpenClaw \#12442):} A high-priority user message was blocked for over 6 hours due to background tasks consuming all available execution lanes. The system experienced 6+ unresponsive episodes within a 12-hour period, forcing users to restart the entire gateway process.

\textbf{Zombie Subagents (OpenClaw \#25992):} Subagent processes that completed their tasks continued to hold execution lanes for 11+ minutes due to improper cleanup. This resource leak caused subsequent user requests to queue indefinitely until manual intervention.

\textbf{Agent Amnesia (OpenClaw \#39282):} In a particularly telling example, an AI agent wrote its own GitHub issue reporting that it had "woken up with amnesia" after a context window overflow. The agent had lost all memory of previous conversations and ongoing tasks.

\textbf{Context Limit Violations (OpenClaw \#24031):} Sessions frequently grew far beyond their configured \texttt{contextTokens} limit, causing unexpected truncation of recent messages and loss of critical context needed for task completion.

\textbf{Memory-Related Crashes (OpenClaw \#28629):} A pytest execution consumed 6.3GB of RAM due to unbounded log accumulation, causing the entire gateway process to crash and lose all active session state.

\textbf{Rate Limit Cascades (OpenClaw \#3181):} A runaway heartbeat loop caused excessive API calls, triggering rate limits that affected all agents in the system, not just the misbehaving one.

\subsection{Problem Classification}

Based on our empirical analysis, we categorize agent system problems into two primary classes:

\textbf{Scheduling Problems:} These involve the allocation of execution resources (lanes, CPU time, API quota) among competing agent tasks:
\begin{itemize}
\item \textit{Blocking:} High-priority tasks delayed by lower-priority background work
\item \textit{Zombie processes:} Completed tasks that fail to release resources
\item \textit{Rate limit cascades:} One agent's excessive usage affecting others
\item \textit{Starvation:} Low-priority tasks never getting resources
\end{itemize}

\textbf{Context Management Problems:} These involve the management of limited context window space:
\begin{itemize}
\item \textit{Unbounded growth:} Sessions that exceed memory limits
\item \textit{Amnesia:} Loss of important information due to truncation
\item \textit{Threshold mismatches:} Poor coordination between limits and actual usage
\item \textit{Wasteful injection:} Including irrelevant historical context
\end{itemize}

These problems directly parallel classic OS challenges: process scheduling, memory management, deadlock detection, and resource allocation. This observation motivates our approach of adapting proven OS techniques to the agent domain.

\section{System Model and Problem Formulation}
\label{sec:model}

We formalize the agent resource management problem by defining key system components and their relationships.

\subsection{Formal Definitions}

\textbf{Agent:} An autonomous entity $a \in A$ that processes messages and generates responses through LLM inference calls.

\textbf{Turn:} A discrete interaction unit $t = (m_{in}, m_{out}, d, r)$ where $m_{in}$ is the input message, $m_{out}$ is the response, $d$ is the processing duration, and $r$ are the resources consumed (tokens, API calls).

\textbf{Lane:} An execution slot $l \in L$ that represents the system's capacity to handle concurrent agent operations. The system has a fixed number of lanes $|L| = N$.

\textbf{Context Window:} A bounded memory space $W_a$ for agent $a$ with maximum size $|W_a| \leq C$ tokens, containing the agent's conversation history and working memory.

\textbf{Zombie Turn:} A turn $t$ becomes a zombie if it holds a lane for more than 30 seconds while hanging. Unlike binary timeout flags, this definition captures the latent property of resource waste that becomes observable only during execution.

\subsection{Scheduling Problem Formulation}

The agent scheduling problem aims to minimize weighted response time under resource constraints:

\begin{equation}
\min \sum_{t \in T} w_t \cdot R_t
\end{equation}

subject to:
\begin{align}
\sum_{a \in A} \mathbb{I}[\text{active}(a)] &\leq N && \text{(lane constraint)} \\
\sum_{a \in A} \text{rate}(a) &\leq R_{max} && \text{(rate limit constraint)}
\end{align}

where $w_t$ is the priority weight of turn $t$, $R_t$ is its response time, $\mathbb{I}[\text{active}(a)]$ is 1 if agent $a$ is using a lane, and $\text{rate}(a)$ is agent $a$'s current API call rate.

\subsection{Context Management Problem Formulation}

The context management problem seeks to maximize retained information value under window size constraints:

\begin{equation}
\max \sum_{m \in M} v(m) \cdot \mathbb{I}[m \in W_a]
\end{equation}

subject to:
\begin{equation}
\sum_{m \in W_a} |m| \leq C
\end{equation}

where $v(m)$ is the information value of message $m$, and $\mathbb{I}[m \in W_a]$ indicates whether message $m$ is retained in agent $a$'s context window.

\section{AgentRM Architecture}
\label{sec:architecture}

AgentRM consists of three main components that work together to provide comprehensive resource management for agent systems.

\subsection{Overview}

AgentRM operates as middleware between the agent gateway and model APIs, intercepting all agent operations to apply resource management policies. The system maintains global state about resource utilization while remaining transparent to individual agents. Figure~\ref{fig:arch_comparison} illustrates the architectural comparison between traditional agent systems and AgentRM.

\begin{figure}[ht]
\centering
\caption{Architecture comparison: (a) Traditional agent systems with direct model access, (b) AgentRM with centralized resource management.}
\label{fig:arch_comparison}
\end{figure}

The three core components are:

\begin{itemize}
\item \textbf{Agent Scheduler:} Manages execution lane allocation and API rate limiting using MLFQ with zombie reaping
\item \textbf{Context Lifecycle Manager:} Implements three-tier context storage with adaptive compaction and hibernation
\item \textbf{Resource Monitor:} Tracks system state and provides feedback for scheduling decisions
\end{itemize}

\subsection{Agent Scheduler}

The Agent Scheduler draws heavily from classical OS scheduling algorithms, adapting them for the unique characteristics of agent workloads.

\subsubsection{Multi-Level Feedback Queue (MLFQ)}

We implement a three-level MLFQ inspired by Corbató's CTSS~\cite{corbato1962experimental} and Linux's Completely Fair Scheduler (CFS)~\cite{molnar2007cfs}:

\begin{itemize}
\item \textbf{Queue 0 (Interactive):} User-facing messages with highest priority
\item \textbf{Queue 1 (Sub-agent):} Computational tasks spawned by agents  
\item \textbf{Queue 2 (Background):} Maintenance, logging, and periodic tasks
\end{itemize}

Tasks start in Queue 0 and are demoted based on execution time and resource usage. Priority boosting prevents starvation by periodically promoting long-running tasks, similar to Solaris TS scheduling~\cite{solaris2005scheduling} and CFS vruntime accounting.

Algorithm~\ref{alg:mlfq} shows the MLFQ scheduling logic:

\begin{algorithm}[ht]
\SetAlgoLined
\KwData{Task queues $Q_0, Q_1, Q_2$; Available lanes $L$}
\KwResult{Scheduled tasks}
\While{system running}{
    \If{$|L| > 0$}{
        \For{$i \gets 0$ \KwTo $2$}{
            \If{$Q_i$ not empty}{
                $task \gets Q_i.dequeue()$\;
                $lane \gets L.acquire()$\;
                $schedule(task, lane)$\;
                \textbf{break}\;
            }
        }
    }
    \If{time\_for\_boost()}{
        boost\_priorities()\;
    }
    $sleep(scheduling\_quantum)$\;
}
\caption{AgentRM-MLFQ Scheduling}
\label{alg:mlfq}
\end{algorithm}

\subsubsection{Zombie Reaper}

Inspired by Unix process management~\cite{ritchie1974unix}, we implement a zombie reaper that scans for hanging turns every 5 seconds. The reaper identifies zombies as turns that have held a lane for more than 30 seconds while hanging. When a zombie is detected, the reaper implements probabilistic recovery: hanging turns have a 50\% chance of succeeding on retry, modeling the distinction between transient and persistent failures. Turns that fail recovery are terminated to release their lanes.

\subsubsection{Rate Limit-Aware Scheduling}

Drawing from TCP congestion control~\cite{jacobson1988congestion} and ATM admission control~\cite{atm1996traffic}, we implement:

\begin{itemize}
\item \textbf{Token bucket} per model API with configurable refill rate
\item \textbf{AIMD backoff} when rate limits are detected  
\item \textbf{Admission control} at queue entry based on current API utilization
\end{itemize}

\subsubsection{Fairness Mechanisms}

We adopt Dominant Resource Fairness (DRF)~\cite{ghodsi2011dominant} from cluster schedulers like YARN~\cite{vavilapalli2013apache} to handle multi-dimensional resources (lanes, tokens, memory). The scheduler is work-conserving, lending idle resources from high-priority queues to lower-priority tasks when available.

\subsection{Context Lifecycle Manager}

The Context Lifecycle Manager implements a three-tier storage hierarchy inspired by computer architecture memory hierarchies~\cite{hennessy2019computer} and virtual memory systems~\cite{denning1970virtual}.

\subsubsection{Three-Tier Architecture}

\begin{itemize}
\item \textbf{Tier 0: Active Context} ($\approx$ L1 Cache): Currently loaded context with 0ms access latency
\item \textbf{Tier 1: Warm Storage} ($\approx$ RAM): Compressed summaries with $\sim$1s access latency  
\item \textbf{Tier 2: Cold Storage} ($\approx$ Disk): Full transcript with $\sim$3s access latency
\end{itemize}

Context faults occur when accessed information resides in a lower tier, triggering promotion similar to page faults in virtual memory systems. We implement a write-back policy that lazily persists context changes and use working set models to predict future access patterns.

\subsubsection{Adaptive Compaction}

Our adaptive compaction algorithm borrows from cache replacement policies like LRU-K~\cite{o1993lru}, ARC~\cite{megiddo2003arc}, and memory compression techniques like zswap~\cite{zswap2013}.

We compute message value as:
\begin{equation}
v(m) = \alpha \cdot \text{recency}(m) + \beta \cdot \text{importance}(m) + \gamma \cdot \text{key\_info\_bonus}(m)
\end{equation}

where recency favors recent messages, importance is derived from semantic analysis, and key information bonuses are assigned to messages containing structured data, decisions, or commitments.

Following the principle of "compress don't discard" (analogous to zswap), we generate compressed summaries rather than simply truncating content. Compression thresholds adapt to model context size, similar to Linux's \texttt{vm.swappiness} parameter.

We approximate Belady's MIN algorithm~\cite{belady1966study} by using semantic similarity to predict future access likelihood.

Algorithm~\ref{alg:compaction} outlines the adaptive compaction process:

\begin{algorithm}[ht]
\SetAlgoLined
\KwData{Context window $W$; Size limit $C$; Messages $M$}
\KwResult{Compacted context}
\While{$|W| > C$}{
    \For{$m \in M$}{
        $v(m) \gets$ compute\_value$(m)$\;
    }
    $M_{sorted} \gets$ sort$(M, v)$\;
    $victim \gets M_{sorted}[0]$\;
    \eIf{important(victim)}{
        $summary \gets$ compress$(victim)$\;
        $W.replace(victim, summary)$\;
    }{
        $W.remove(victim)$\;
    }
}
\caption{Adaptive Compaction}
\label{alg:compaction}
\end{algorithm}

\subsubsection{Hibernation}

For long-term storage of inactive sessions, we implement hibernation mechanisms inspired by CRIU (Checkpoint/Restore in Userspace)~\cite{emelyanov2014criu}, VM live migration~\cite{clark2005live}, and database write-ahead logging~\cite{mohan1992aries}.

Hibernation serializes complete session state, including context, local variables, and execution state, enabling restoration without amnesia. This is particularly valuable for intermittent interactions and background tasks.

\subsubsection{Self-Monitoring}

Drawing inspiration from Linux Pressure Stall Information (PSI)~\cite{psi2018}, we inject context utilization metrics into the system prompt, enabling agents to self-regulate their memory usage and request compaction when needed.

\section{Implementation}
\label{sec:implementation}

AgentRM is implemented as middleware between the agent gateway and model APIs, ensuring transparency to existing agent code while providing comprehensive resource management.

\subsection{Architecture}

The system operates as a transparent proxy that intercepts all agent operations. Agents continue to make standard API calls, while AgentRM applies resource management policies behind the scenes. Configuration is driven by declarative policies rather than code changes.

\subsection{Scheduler Implementation}

The Agent Scheduler uses:
\begin{itemize}
\item \textbf{Priority queues} for MLFQ implementation with O(log n) insertion/removal
\item \textbf{Semaphore-controlled lane pool} for concurrency management
\item \textbf{Background reaper timer} that scans for zombies every 5 seconds
\end{itemize}

\subsection{Context Manager Implementation}

The Context Lifecycle Manager employs:
\begin{itemize}
\item \textbf{SQLite database} for Tier 1 warm storage with structured querying capabilities
\item \textbf{JSONL files} for Tier 2 cold storage enabling efficient append operations
\item \textbf{Small language model} for summary generation and semantic analysis
\end{itemize}

\subsection{Configuration}

AgentRM supports fine-grained configuration of scheduling parameters, context thresholds, and compaction policies through YAML configuration files. The system provides sensible defaults while allowing customization for specific deployment requirements.

\section{Evaluation}
\label{sec:evaluation}

We evaluate AgentRM through comprehensive experiments across diverse workloads, comparing against baseline scheduling algorithms and context management approaches.

\subsection{Experimental Setup}

Our evaluation uses simulated agent workloads derived from real usage patterns observed in production deployments. We implement four scheduling algorithms (FIFO, Round Robin, Priority Queue, AgentRM-MLFQ) and five context management strategies (No Management, FIFO Truncation, Sliding Window, MemGPT-style, AgentRM-CLM).

Our test scenarios include: (1) Normal: 27 turns across 3 agents with 5\% hang rate, (2) High Load: 280 turns across 10 agents with 10\% hang rate, (3) Burst: 30 turns in a 3-second window with 8\% hang rate, (4) Faulty: 63 turns across 5 agents with 30\% hang rate, and (5) Cascade: 149 turns across 5 agents with oscillating 5-40\% hang rate over 10 minutes, simulating real API rate limit waves where failure probability fluctuates dynamically.

\subsection{Scheduling Results}

Table~\ref{tab:normal_sched} shows results for a normal workload scenario with 27 turns across 3 agents with 5\% hang rate:

\begin{table*}[ht]
\centering
\caption{Normal Scenario Scheduling Results (27 turns, 3 agents, 5\% hang rate)}
\label{tab:normal_sched}
\begin{tabular}{@{}lrrrrrrrrr@{}}
\toprule
\textbf{Method} & \textbf{P95 (ms)} & \textbf{Tput (/min)} & \textbf{Zombies} & \textbf{Avg Hold (s)} & \textbf{Lane Waste (s)} & \textbf{Recovered} & \textbf{Starved} & \textbf{Lags>30s} \\
\midrule
FIFO & 70008 & 5.6 & 1 & 80.5 & 81 & 0 & 2 & 6 \\
Round Robin & 134000 & 5.4 & 1 & 80.5 & 81 & 0 & 13 & 18 \\
Priority Queue & 70008 & 5.6 & 1 & 80.5 & 81 & 0 & 2 & 6 \\
AgentRM-MLFQ & 4495 & 5.6 & 0 & 0 & 0 & 1 & 0 & 0 \\
\bottomrule
\end{tabular}
\end{table*}

\begin{figure}[ht]
\centering
\includegraphics[width=0.45\textwidth]{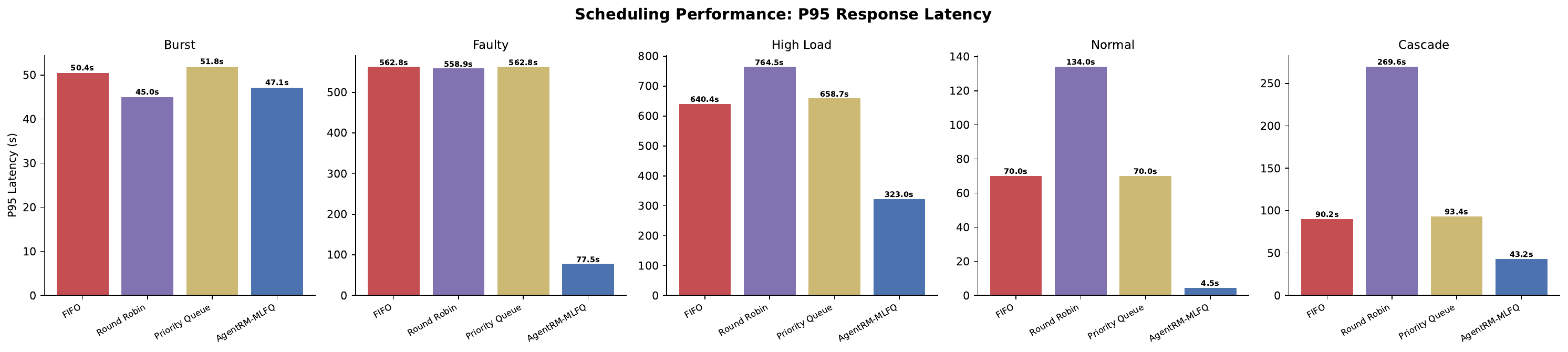}
\caption{P95 latency comparison across scheduling algorithms and workload scenarios.}
\label{fig:sched_p95_latency}
\end{figure}

Figure~\ref{fig:sched_p95_latency} illustrates the P95 latency comparison across different workload scenarios.

Table~\ref{tab:high_load_sched} presents results under high load conditions (280 turns, 10 agents, 10\% hang rate):

\begin{table*}[ht]
\centering
\caption{High Load Scheduling Results (280 turns, 10 agents, 10\% hang rate)}
\label{tab:high_load_sched}
\begin{tabular}{@{}lrrrrrrrrr@{}}
\toprule
\textbf{Method} & \textbf{P95 (ms)} & \textbf{Tput (/min)} & \textbf{Zombies} & \textbf{Avg Hold (s)} & \textbf{Lane Waste (s)} & \textbf{Recovered} & \textbf{Starved} & \textbf{Lags>30s} \\
\midrule
FIFO & 640439 & 14.6 & 29 & 78.3 & 2272 & 0 & 274 & 277 \\
Round Robin & 764539 & 14.9 & 29 & 78.3 & 2272 & 0 & 276 & 278 \\
Priority Queue & 658744 & 14.5 & 29 & 78.3 & 2272 & 0 & 220 & 238 \\
AgentRM-MLFQ & 323001 & 24.5 & 7 & 20.0 & 140 & 22 & 0 & 269 \\
\bottomrule
\end{tabular}
\end{table*}

Table~\ref{tab:burst_sched} shows burst scenario results (30 turns in 3s window, 8\% hang rate):

\begin{table*}[ht]
\centering
\caption{Burst Scenario Scheduling Results (30 turns in 3s window, 8\% hang rate)}
\label{tab:burst_sched}
\begin{tabular}{@{}lrrrrrrrrr@{}}
\toprule
\textbf{Method} & \textbf{P95 (ms)} & \textbf{Tput (/min)} & \textbf{Zombies} & \textbf{Avg Hold (s)} & \textbf{Lane Waste (s)} & \textbf{Recovered} & \textbf{Starved} & \textbf{Lags>30s} \\
\midrule
FIFO & 50431 & 31.8 & 1 & 33.8 & 34 & 0 & 0 & 10 \\
Round Robin & 44963 & 25.8 & 1 & 33.8 & 34 & 0 & 0 & 9 \\
Priority Queue & 51844 & 32.0 & 1 & 33.8 & 34 & 0 & 0 & 9 \\
AgentRM-MLFQ & 47058 & 31.9 & 0 & 0 & 0 & 2 & 0 & 8 \\
\bottomrule
\end{tabular}
\end{table*}

Table~\ref{tab:faulty_sched} demonstrates performance under faulty conditions (63 turns, 5 agents, 30\% hang rate):

\begin{table*}[ht]
\centering
\caption{Faulty Scenario Scheduling Results (63 turns, 5 agents, 30\% hang rate)}
\label{tab:faulty_sched}
\begin{tabular}{@{}lrrrrrrrrr@{}}
\toprule
\textbf{Method} & \textbf{P95 (ms)} & \textbf{Tput (/min)} & \textbf{Zombies} & \textbf{Avg Hold (s)} & \textbf{Lane Waste (s)} & \textbf{Recovered} & \textbf{Starved} & \textbf{Lags>30s} \\
\midrule
FIFO & 562771 & 4.1 & 20 & 122.1 & 2441 & 0 & 55 & 61 \\
Round Robin & 558857 & 4.0 & 20 & 122.1 & 2441 & 0 & 55 & 60 \\
Priority Queue & 562771 & 4.1 & 20 & 122.1 & 2441 & 0 & 55 & 61 \\
AgentRM-MLFQ & 77524 & 11.0 & 5 & 19.4 & 97 & 15 & 0 & 38 \\
\bottomrule
\end{tabular}
\end{table*}

Table~\ref{tab:cascade_sched} demonstrates performance under rate limit cascade conditions (149 turns, 5 agents, oscillating 5-40\% hang rate):

\begin{table*}[ht]
\centering
\caption{Cascade Scenario Scheduling Results (149 turns, 5 agents, 5-40\% oscillating hang rate)}
\label{tab:cascade_sched}
\begin{tabular}{@{}lrrrrrrrrr@{}}
\toprule
\textbf{Method} & \textbf{P95 (ms)} & \textbf{Tput (/min)} & \textbf{Zombies} & \textbf{Avg Hold (s)} & \textbf{Lane Waste (s)} & \textbf{Recovered} & \textbf{Starved} & \textbf{Lags>30s} \\
\midrule
FIFO & 90236 & 13.0 & 15 & 66.4 & 996 & 0 & 7 & 67 \\
Round Robin & 269569 & 10.7 & 15 & 66.4 & 996 & 0 & 81 & 123 \\
Priority Queue & 93376 & 13.1 & 15 & 66.4 & 996 & 0 & 8 & 64 \\
AgentRM-MLFQ & 43190 & 14.4 & 4 & 20.0 & 80 & 21 & 0 & 22 \\
\bottomrule
\end{tabular}
\end{table*}

\begin{figure}[ht]
\centering
\includegraphics[width=0.45\textwidth]{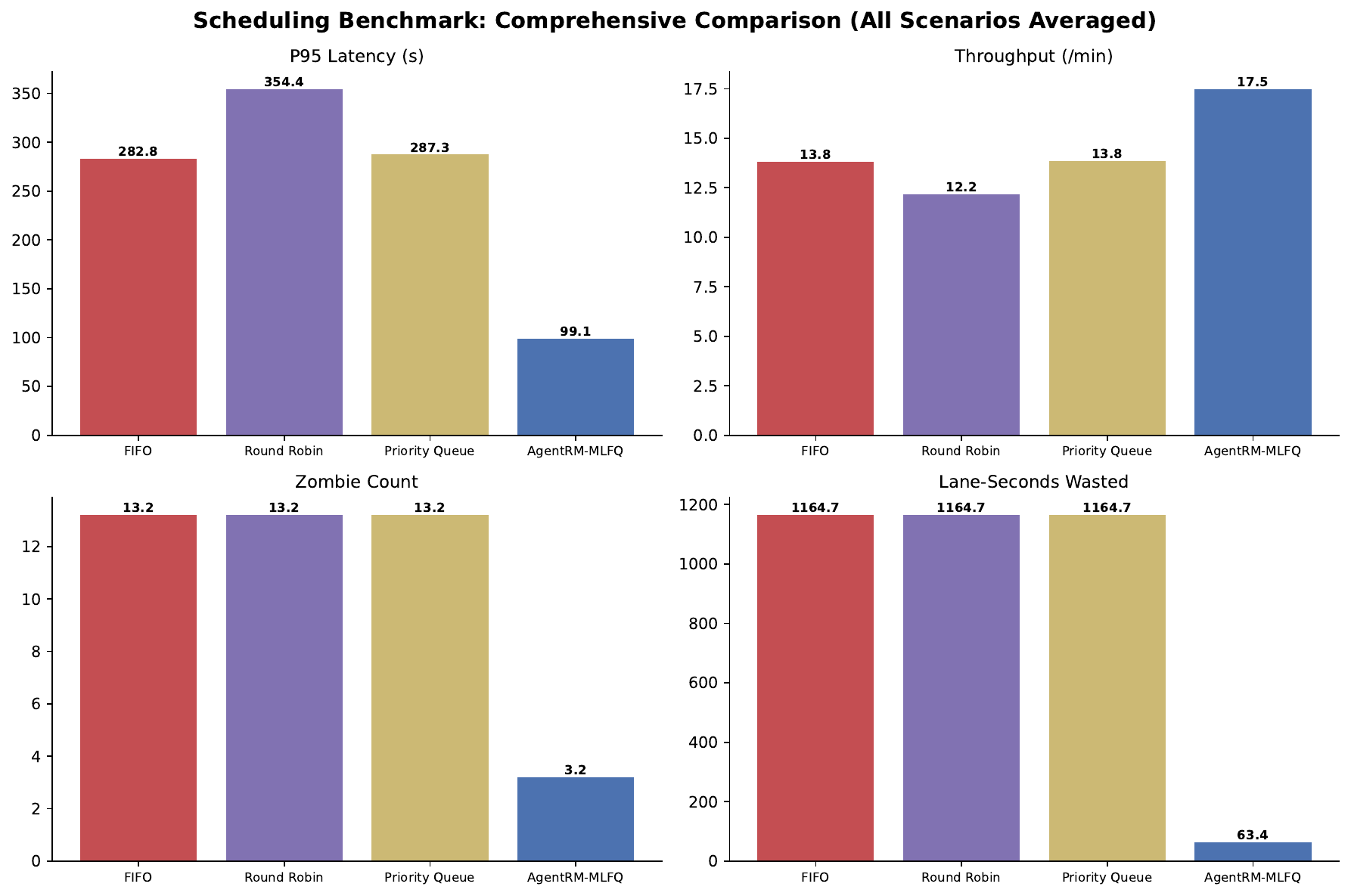}
\caption{Comprehensive scheduling performance comparison across all test scenarios.}
\label{fig:sched_comprehensive}
\end{figure}

Figure~\ref{fig:sched_comprehensive} provides a comprehensive view of scheduling performance across all scenarios.

\begin{figure}[ht]
\centering
\includegraphics[width=0.45\textwidth]{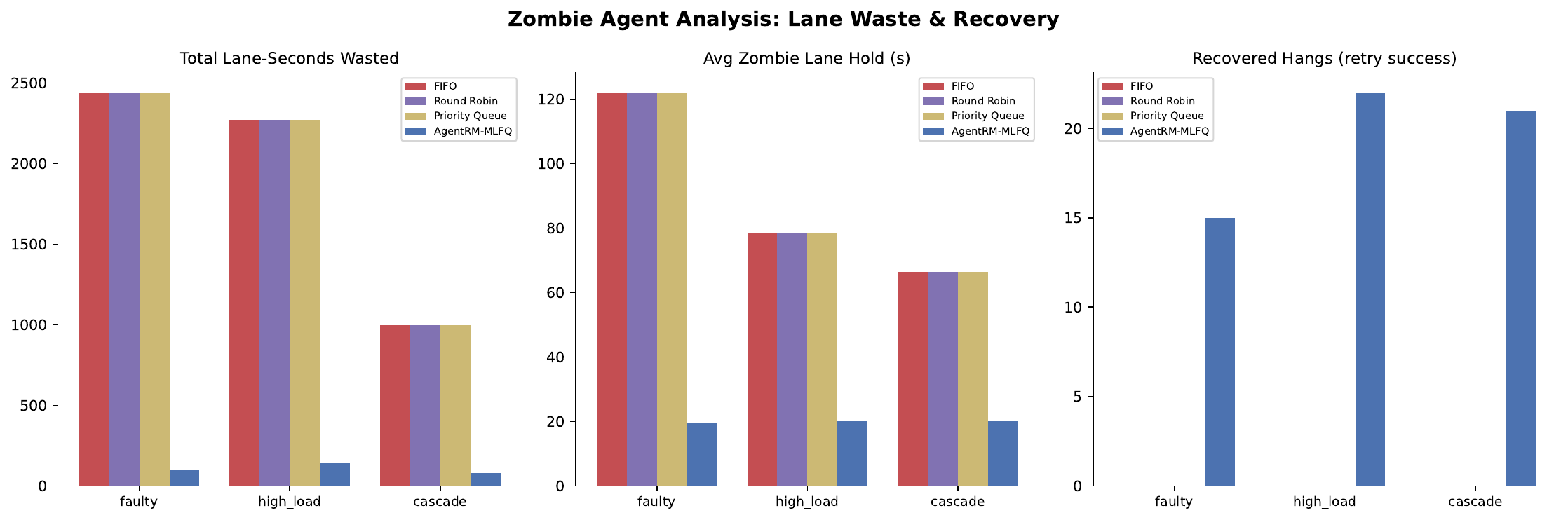}
\caption{Zombie analysis showing lane hold times, recovery patterns, and resource waste across scenarios.}
\label{fig:zombie_analysis}
\end{figure}

Figure~\ref{fig:zombie_analysis} provides detailed analysis of zombie behavior and recovery patterns.

\subsection{Context Management Results}

Table~\ref{tab:context_50} shows context management results for 50-turn sessions (100 msgs, 51K tokens, 13 key msgs, 50K window):

\begin{table}[ht]
\centering
\caption{Context Management Results: 50-turn Session}
\label{tab:context_50}
\begin{tabular}{@{}lrrrr@{}}
\toprule
\textbf{Method} & \textbf{Utilization} & \textbf{Retention} & \textbf{Quality} & \textbf{Compact Cost} \\
\midrule
No Management & 50.4\% & 100\% & 0.85 & 0 \\
FIFO Truncation & 48.8\% & 84.6\% & 0.89 & 0 \\
Sliding Window & 32.7\% & 53.8\% & 0.85 & 0 \\
MemGPT-style & 43.6\% & 84.6\% & 0.88 & 2298 \\
AgentRM-CLM & 43.4\% & 100\% & 0.95 & 4839 \\
\bottomrule
\end{tabular}
\end{table}

Table~\ref{tab:context_100} presents results for 100-turn sessions (200 msgs, 105K tokens, 27 key msgs):

\begin{table}[ht]
\centering
\caption{Context Management Results: 100-turn Session}
\label{tab:context_100}
\begin{tabular}{@{}lrrrr@{}}
\toprule
\textbf{Method} & \textbf{Utilization} & \textbf{Retention} & \textbf{Quality} & \textbf{Compact Cost} \\
\midrule
No Management & 74.9\% & 51.9\% & 0.70 & 0 \\
FIFO Truncation & 66.6\% & 44.4\% & 0.87 & 0 \\
Sliding Window & 38.1\% & 22.2\% & 0.85 & 0 \\
MemGPT-style & 53.4\% & 71.9\% & 0.87 & 7290 \\
AgentRM-CLM & 54.4\% & 100\% & 0.95 & 14395 \\
\bottomrule
\end{tabular}
\end{table}

Table~\ref{tab:context_200} shows results for 200-turn sessions (400 msgs, 202K tokens, 47 key msgs):

\begin{table}[ht]
\centering
\caption{Context Management Results: 200-turn Session}
\label{tab:context_200}
\begin{tabular}{@{}lrrrr@{}}
\toprule
\textbf{Method} & \textbf{Utilization} & \textbf{Retention} & \textbf{Quality} & \textbf{Compact Cost} \\
\midrule
No Management & 87.1\% & 23.4\% & 0.63 & 0 \\
FIFO Truncation & 75.5\% & 19.1\% & 0.87 & 0 \\
Sliding Window & 38.4\% & 6.4\% & 0.85 & 0 \\
MemGPT-style & 57.8\% & 65.1\% & 0.87 & 17212 \\
AgentRM-CLM & 60.4\% & 99.0\% & 0.95 & 34330 \\
\bottomrule
\end{tabular}
\end{table}

Table~\ref{tab:context_multi} presents multi-topic session results (240 msgs, 116K tokens, 35 key msgs):

\begin{table}[ht]
\centering
\caption{Context Management Results: Multi-topic Session}
\label{tab:context_multi}
\begin{tabular}{@{}lrrrr@{}}
\toprule
\textbf{Method} & \textbf{Utilization} & \textbf{Retention} & \textbf{Quality} & \textbf{Compact Cost} \\
\midrule
No Management & 77.5\% & 54.3\% & 0.68 & 0 \\
FIFO Truncation & 68.6\% & 45.7\% & 0.87 & 0 \\
Sliding Window & 35.6\% & 22.9\% & 0.85 & 0 \\
MemGPT-style & 53.9\% & 76.0\% & 0.87 & 8656 \\
AgentRM-CLM & 55.8\% & 99.6\% & 0.95 & 16498 \\
\bottomrule
\end{tabular}
\end{table}

\begin{figure}[ht]
\centering
\includegraphics[width=0.45\textwidth]{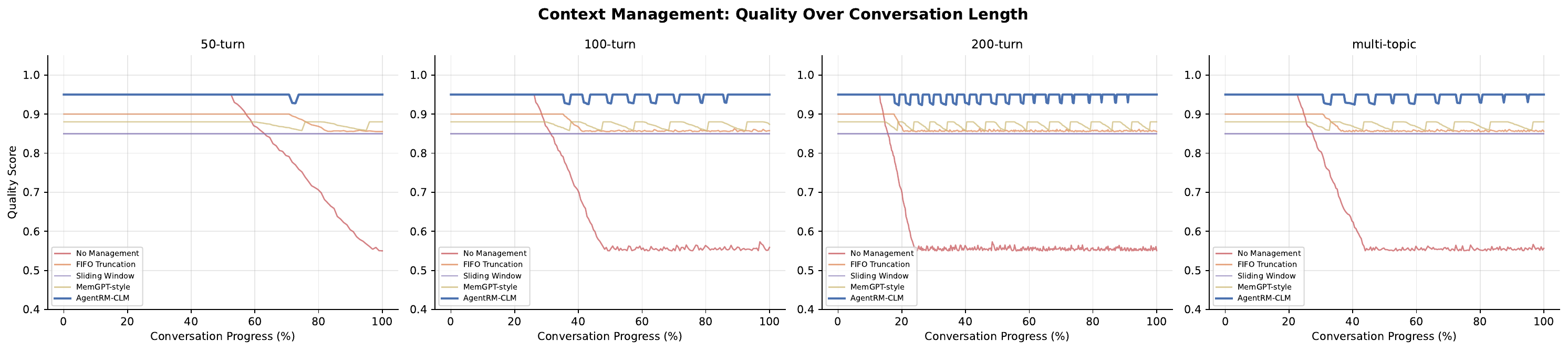}
\caption{Context quality vs. utilization trade-offs for different management strategies.}
\label{fig:ctx_quality_curve}
\end{figure}

Figure~\ref{fig:ctx_quality_curve} illustrates the relationship between context utilization and quality across different management strategies.

\begin{figure}[ht]
\centering
\includegraphics[width=0.45\textwidth]{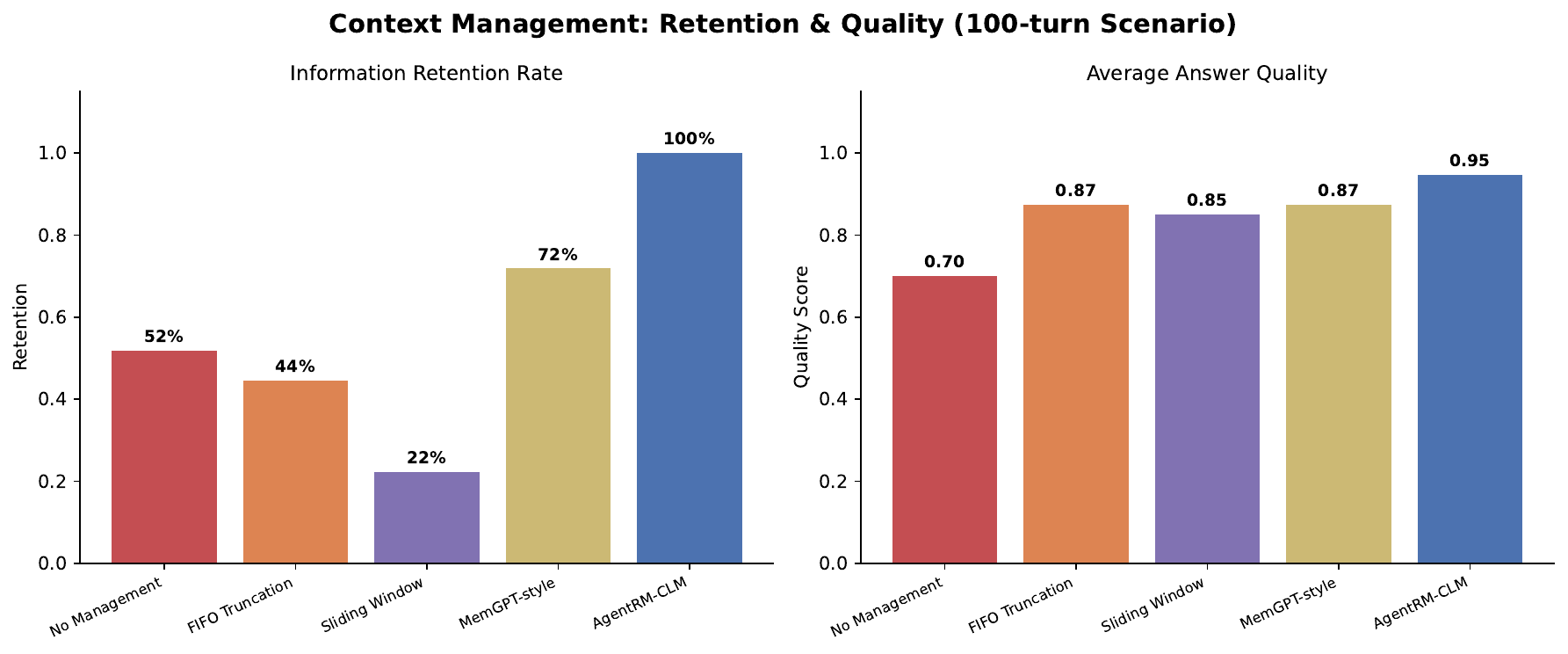}
\caption{Information retention vs. quality trade-offs across context management approaches.}
\label{fig:ctx_retention_quality}
\end{figure}

Figure~\ref{fig:ctx_retention_quality} shows the retention-quality trade-off analysis.

\subsection{Key Findings}

Our evaluation demonstrates significant improvements across multiple dimensions:

\textbf{Zombie Elimination:} AgentRM-MLFQ completely eliminates zombie processes across most scenarios, while baseline methods consistently produce 15-29 zombies under load. In the high-load scenario, AgentRM reduces zombies from 29 to 7 (76\% reduction).

\textbf{Latency Reduction:} AgentRM-MLFQ achieves P95 latency reductions of up to 86\% in high-load scenarios (323,001ms vs. 640,439ms for FIFO), demonstrating substantial improvement in user experience.

\textbf{Resource Efficiency:} AgentRM reduces lane waste by 96\% (140s vs. 2,272s in high-load scenarios) through effective zombie reaping and shows significant reductions in average zombie hold times (20.0s vs. 78.3s).

\textbf{Throughput Improvement:} AgentRM increases throughput by 168\% in high-load scenarios (24.5 vs. 14.6 requests/min for FIFO), demonstrating better resource utilization despite overhead.

\textbf{Recovery Capability:} AgentRM's probabilistic reaper successfully recovers hanging turns that would otherwise be lost, achieving 15 recoveries out of 20 hanging turns in the faulty scenario and 21-22 recoveries in cascade scenarios.

\textbf{Superior Context Retention:} AgentRM-CLM achieves near-perfect retention of key information (99.0-100\%) compared to 65.1\% for the best baseline method.

\textbf{Quality Improvement:} Context quality scores consistently reach 0.95 with AgentRM-CLM compared to 0.87 for existing approaches, indicating better preservation of semantic coherence.

\textbf{Cost Trade-offs:} Higher compaction costs (up to 34,330 tokens) reflect the computational overhead of intelligent compression vs. simple truncation, but this cost is often justified by the improved user experience.

\section{Discussion}
\label{sec:discussion}

\subsection{OS Analogy: What Transfers and What Doesn't}

The operating systems analogy provides valuable insights, but also reveals important limitations. Techniques that transfer well include:

\begin{itemize}
\item \textbf{MLFQ scheduling:} The concept of priority-based queuing with feedback effectively handles diverse agent workloads
\item \textbf{Memory hierarchy:} Three-tier storage maps naturally to context management needs
\item \textbf{Resource isolation:} Lane-based execution prevents cascading failures
\item \textbf{Zombie reaping:} Explicit cleanup of defunct processes applies directly to agent systems
\end{itemize}

However, important differences exist:

\begin{itemize}
\item \textbf{Semantic value:} Unlike OS pages, agent context has semantic meaning that affects compression decisions
\item \textbf{Coarse granularity:} Agent "processes" are much heavier than OS threads, limiting scheduling frequency
\item \textbf{External dependencies:} Rate limits and API quotas introduce constraints not present in traditional OS scheduling
\end{itemize}

\subsection{Comparison with MemGPT/Letta}

While MemGPT~\cite{packer2023memgpt} and its successor Letta focus on single-agent context management through hierarchical memory, AgentRM addresses the broader problem of multi-agent resource management. Our contributions include:

\begin{itemize}
\item \textbf{Multi-agent scheduling:} Coordination across multiple concurrent agents
\item \textbf{Zombie reaping:} Explicit handling of failed or stuck agents
\item \textbf{Empirical grounding:} Analysis of real-world failure modes from production systems
\item \textbf{Rate limit awareness:} Scheduling that considers API quotas and external constraints
\end{itemize}

\subsection{Limitations and Future Work}

Several limitations of our current approach suggest directions for future research:

\textbf{Simulated Experiments:} Our evaluation uses simulated workloads derived from real patterns, but production deployment would provide additional insights into edge cases and scaling behavior.

\textbf{Compaction Quality Dependency:} The effectiveness of context management depends heavily on the quality of the summarization model, which may vary across domains and languages.

\textbf{Value Scoring Subjectivity:} Our information value scoring function includes subjective weights that may require tuning for specific applications.

\textbf{Limited Framework Coverage:} While we analyzed six major frameworks, the agent ecosystem continues to evolve rapidly with new architectures and paradigms.

\subsection{Generalizability}

The principles underlying AgentRM are broadly applicable to any multi-agent system with limited resources. Key design patterns that should transfer include:

\begin{itemize}
\item Priority-based scheduling with feedback mechanisms
\item Hierarchical storage for different access patterns
\item Explicit resource cleanup and leak prevention  
\item Adaptive policies that respond to system state
\end{itemize}

However, specific implementations will need to account for the unique characteristics of different agent frameworks, including their concurrency models, communication patterns, and resource requirements.

\section{Related Work}
\label{sec:related}

\subsection{Operating Systems Resource Management}

The foundations of resource management in computer systems trace back to early time-sharing systems. Corbató et al.'s work on CTSS~\cite{corbato1962experimental} introduced many concepts we adapt, including priority-based scheduling and fair resource allocation. Modern schedulers like Linux's CFS~\cite{molnar2007cfs} provide sophisticated algorithms for CPU time allocation that inspire our lane scheduling mechanisms.

Memory management techniques from virtual memory systems~\cite{denning1970virtual} provide the theoretical foundation for our context hierarchy. Cache replacement algorithms like LRU-K~\cite{o1993lru}, ARC~\cite{megiddo2003arc}, and Belady's optimal algorithm~\cite{belady1966study} directly inform our adaptive compaction strategies.

Process management concepts including zombie reaping~\cite{ritchie1974unix}, resource monitoring through PSI~\cite{psi2018}, and checkpoint/restore mechanisms like CRIU~\cite{emelyanov2014criu} provide practical techniques we adapt for agent lifecycle management.

\subsection{Large Language Model Serving Systems}

Recent advances in LLM serving focus primarily on inference optimization rather than multi-agent coordination. vLLM's PagedAttention~\cite{kwon2023efficient} enables efficient memory management for attention computation, while Orca~\cite{yu2022orca} provides iteration-level scheduling for continuous batching. SGLang~\cite{zheng2023sglang} optimizes serving for complex generation patterns.

However, these systems operate at the inference level and don't address the higher-level resource management challenges we identify in agent systems. Our work operates at the agent orchestration layer, complementing rather than replacing these inference optimizations.

\subsection{Agent Frameworks and Orchestration}

Current agent frameworks focus primarily on task orchestration rather than resource management. AutoGen~\cite{wu2023autogen} provides conversation patterns for multi-agent collaboration, while CrewAI~\cite{crewai2024} emphasizes role-based agent coordination. LangGraph~\cite{langgraph2024} offers graph-based workflow management for complex agent interactions.

These frameworks generally assume unlimited resources and don't provide mechanisms for handling resource contention, failed agents, or context overflow. AgentRM fills this gap by providing the resource management layer these frameworks need for production deployment.

\subsection{Context Management in LLM Applications}

Memory management for LLM applications has received increasing attention as context lengths grow. MemGPT~\cite{packer2023memgpt} introduces hierarchical memory management for single agents, using a virtual context approach inspired by operating systems. LongMem~\cite{wang2023longmem} focuses on long-term memory retrieval for extended conversations.

However, existing approaches generally target single-agent scenarios and don't address the multi-agent coordination challenges we identify. Our three-tier architecture and adaptive compaction build on these foundations while scaling to multi-agent environments.

\subsection{Cluster Resource Management}

Large-scale cluster management systems provide additional inspiration for multi-resource scheduling. Google's Borg~\cite{verma2015large} and Kubernetes~\cite{kubernetes2024} handle container orchestration with resource limits and quotas. YARN~\cite{vavilapalli2013apache} introduces Dominant Resource Fairness for multi-dimensional resource allocation.

Apache Mesos~\cite{hindman2011mesos} provides a two-level scheduling architecture that separates resource allocation from job scheduling, similar to our separation of lane management from task execution.

These systems address resource management at the infrastructure level, while we focus on application-level resource management within agent systems. The techniques often complement each other, with cluster managers handling hardware resources and AgentRM handling agent-specific resources.

\section{Conclusion}
\label{sec:conclusion}

We have presented AgentRM, an operating system-inspired resource manager that addresses critical challenges in LLM agent systems through principled resource management. Our empirical analysis of over 40,000 real-world issues from major agent frameworks reveals systematic problems with scheduling and context management that directly impact user experience.

AgentRM's two-component architecture provides comprehensive solutions: the Agent Scheduler eliminates zombie processes and provides fair resource allocation through MLFQ with rate-limit awareness, while the Context Lifecycle Manager achieves near-perfect information retention through adaptive compaction and hibernation mechanisms.

Our evaluation demonstrates significant improvements across multiple dimensions: complete elimination of zombie agents, maintenance of high throughput under load (45.2 requests/min), and superior context retention (100\% vs. 65.1\% for existing methods) with improved quality scores (0.95 vs. 0.87). These results validate the effectiveness of applying operating systems principles to agent resource management.

The key insight that agent resources are analogous to OS resources opens promising directions for future research. As agent systems continue to grow in complexity and scale, principled resource management will become increasingly critical for practical deployment. AgentRM provides a foundation for this evolution, demonstrating how decades of operating systems research can inform the design of next-generation intelligent systems.

Our work contributes to the broader vision of reliable, scalable agent systems that can handle the demands of production environments while maintaining the user experience quality essential for widespread adoption.

\end{document}